# Engineering Digital Systems for Humanity: a Research Roadmap

MARCO AUTILI, University of L'Aquila, Italy
MARTINA DE SANCTIS, PAOLA INVERARDI, and PATRIZIO PELLICCIONE, Gran Sasso Science Institute (GSSI), L'Aquila, Italy

As testified by new regulations like the European AI Act, worries about the human and societal impact of (autonomous) software technologies are becoming of public concern. Human, societal, and environmental values, alongside traditional software quality, are increasingly recognized as essential for sustainability and long-term well-being. Traditionally, systems are engineered taking into account business goals and technology drivers. Considering the growing awareness in the community, in this paper, we argue that engineering of systems should also consider human, societal, and environmental drivers. Then, we identify the macro and technological challenges by focusing on humans and their role while co-existing with digital systems. The first challenge considers humans in a proactive role when interacting with digital systems, i.e., taking initiative in making things happen instead of reacting to events. The second concerns humans having a reactive role in interacting with digital systems, i.e., humans interacting with digital systems as a reaction to events. The third challenge focuses on humans with a passive role, i.e., they experience, enjoy or even suffer the decisions and/or actions of digital systems. The fourth challenge concerns the duality of trust and trustworthiness, with humans playing any role. Building on the new human, societal, and environmental drivers and the macro and technological challenges, we identify a research roadmap of digital systems for humanity. The research roadmap is concretized in a number of research directions organized into four groups: development process, requirements engineering, software architecture and design, and verification and validation.

## 1 INTRODUCTION

Digital systems are increasingly an integral part of our daily lives. They are pervasive and ubiquitous, automating our homes, the transportation we use, and supporting our work and many other dimensions of our life with varying degrees of automation [96].

Consequently, there is a growing need for assurances of correct behavior to prevent undesired consequences. The potential risks related to software are very clear in the case of critical systems, where the criticality can be related to safety, security, business (e.g., software controlling the stocks

Authors' addresses: Marco Autili, marco.autili@univaq.it, University of L'Aquila, Via Vetoio, 1, L'Aquila, Italy, 67100; Martina De Sanctis, martina.desanctis@gssi.it; Paola Inverardi, paola.inverardi@gssi.it; Patrizio Pelliccione, patrizio.pelliccione@gssi.it, Gran Sasso Science Institute (GSSI), L'Aquila, Viale F. Crispi, 7, L'Aquila, Italy, 67100.

trading), or mission (e.g., software controlling satellites). However, nowadays, risks also come from software that is not so obviously considered critical. This is the case for software automating documentation writing, personal assistants, news dissemination support, social network bots, human resources support, investment assistance, etc. AI technologies are key enablers for many, though not all, of these systems. Indeed, the development and popularity of AI, including ML, generative AI, and large language models, triggered significant advancements in unthinkable ways only a few years ago. The growing importance of technology is also testified by the increasing power that technological companies have in spheres that go beyond their business. For instance, controlling communication on Earth or space brings enormous political power, which so far has been reserved for countries and their governments.

It is then becoming clearer that technology and software can have various and impactful consequences on our lives. Human, societal, and environmental (HSE) values, besides the traditional software behavior and quality, are increasingly recognized as important for sustainability and long-term well-being [95]. This pertains to the vast amount of information digital systems collect about us, enabling insights into nearly every aspect of our lives [65, 96]. These insights go beyond privacy concerns, as far as predictions of our future choices and decisions are concerned. The European Commission also prioritizes sustainability, as reflected in its published agenda [44], which outlines the Sustainable Development Goals (SDGs) [118] aimed at fostering smart, sustainable, and inclusive growth. However, sustainability encompasses multiple dimensions – environmental, economic, societal, technical, and individual [17] – while also addressing human moral values and overall well-being. The pervasiveness of digitalization may also present challenges and hinder progress toward achieving the United Nations SDGs, highlighting the need for organizations and companies to deepen their sustainability skills and competencies [53]. The importance of the topic is also testified by the establishment of various regulations, starting from *GDPR* in 2018 and extending to the recent *AI Act* – the first law in the world to regulate the use of AI – recently approved and enacted by the European Parliament [77].

In this paper, we discuss the risks and dangers for people, society, and environemnt (HSE values) with the ambition of identifying a roadmap (a set of research directions) for software engineers. They are involved in the engineering of systems that, we emphasize, should be built for humanity. We should consider that the role of humans in 2030 remains unclear. On the one side, AI will dramatically change the way developers interact with systems [98]. On the other side, the boundaries between developers and end-users will become blurred, as we will discuss in the following, with humans taking a proactive role in the interaction with digital systems. However, this article focuses on human, social, and environmental values that are affected or challenged by digital systems. This implies that pure software engineering challenges that do not affect HSE values, e.g., challenges of requirement engineers, testers, designers, and developers caused by the introduction and use of LLMs, are outside of the scope of this paper.

We start by identifying the HSE drivers that will influence systems engineering together with the traditional business goals and technology drivers. Acknowledging that it is important to consider responsible practices for software developers, in this paper, we focus on HSE drivers from the user perspective and then we discuss how they impact the software engineering activities. Taking into account the HSE drivers, we identify the macro and technological challenges. Our specific perspective in identifying the challenges and the research roadmap is, therefore, to focus on humans and their role and/or relation in their co-existence with digital systems.

The findings of this paper are achieved through a design science research methodology. We performed three iterations. During these iterations, we gathered information from multiple sources: a review of the literature, focus groups conducted during a two-day workshop co-located with FSE 2024, as summarized in [98], an analysis of existing laws and regulations in the field, and insights gained from research projects focused on the ethical aspects of autonomous and robotic systems. In these projects, we had, and still have, the opportunity to collaborate and interact with a multidisciplinary and diverse group of researchers from fields such as philosophy, psychology, sociology, and law, in addition to computer science. More details of the reseach methodology are provided in Section 2.

The first challenge we identify focuses on humans adopting a proactive role when interacting with digital systems, i.e., taking the initiative to make things happen instead of reacting to events that happen. This relates to humans being able to *continuously program systems*, i.e., to program systems at design time but also being able to alter, reprogram, or change the system behavior during runtime. This calls for languages and ways for programming systems (e.g., not necessarily by writing code, but also by examples, through feedback, or voice) accessible to everyone regardless of their background or knowledge. The second challenge concerns humans having a reactive role in the interaction with digital systems. In this challenge, we focus on the "reactive" aspect of human interaction, i.e., on humans that interact with digital systems as a reaction to events. For instance, a real-time monitoring system that prompts engineers to detect issues in a timely manner and provide immediate feedback or take a course of action. The third challenge focuses on humans who have a passive role but experience, enjoy, or suffer, in the worst cases, the decisions and/or actions of digital systems. For instance, this is the case of AI-enabled systems supporting human resources (HR) offices and banking systems in the decision to grant a loan or insurance, and so on. Indeed, in large and complex systems, humans can have more than one role and can be proactive for some functionalities, reactive for some other, and passive for other ones. However, this distinction helps us to better identify the challenges. The fourth transversal challenge concerns the duality of trustworthiness and trust [75]. Trustworthiness concerns the designing of systems to behave safely and guarantee security or quality aspects. Modern AI-enabled systems challenge existing techniques due to the inherent complexity of formulating correct specifications and devising appropriate adequacy metrics to evaluate their effectiveness [91]. Instead, trust concerns the acceptability of systems from the point of view of humans: humans can easily reject algorithmic prescriptions that "authoritatively" make decisions or predictions. For instance, transparency and explainability are certainly qualities that can help establish a trust relationship between humans and machines.

The paper closes by suggesting a research and technological roadmap to engineer systems for humanity. The research roadmap is meant to be for software engineers who aim to build and develop digital systems for humanity. However, some points of the roadmap might also be valuable to anyone interested in other aspects or in the broader impact of digital systems for humanity. Therefore, we organize the research roadmap in research directions concerning the development process, requirements engineering, software architecture and design, and verification and validation.

**Paper structure**: Section 2 describes the research methodology we followed to perform this study. Section 3 motivates the need for HSE drivers, beyond business goals and technology drivers, when engineering systems. Before presenting the 6 HSE drivers, the section introduces some motivation examples and analyses the roles humans can have in digital systems. Section 4 identifies and presents the 4 macro and technological challenges together with their subchallenges. Section 5 builds on the HSE drivers and the macro and technological challenges and presents a research roadmap composed of 14 research directions, which are organized into 4 groups. Finally, the paper concludes with final remarks in Section 6.

## 2 METHODOLOGY

To build a research roadmap on engineering digital systems for humanity, we follow a *design science* [15, 57, 127] research methodology, which is summarized in Figure 1. We perform three iterations, each including the three phases of creating awareness about the problem domain and possible solutions, synthesizing a solution, and validating whether the solution mitigates the problem. In the first iteration, we build awareness based on:

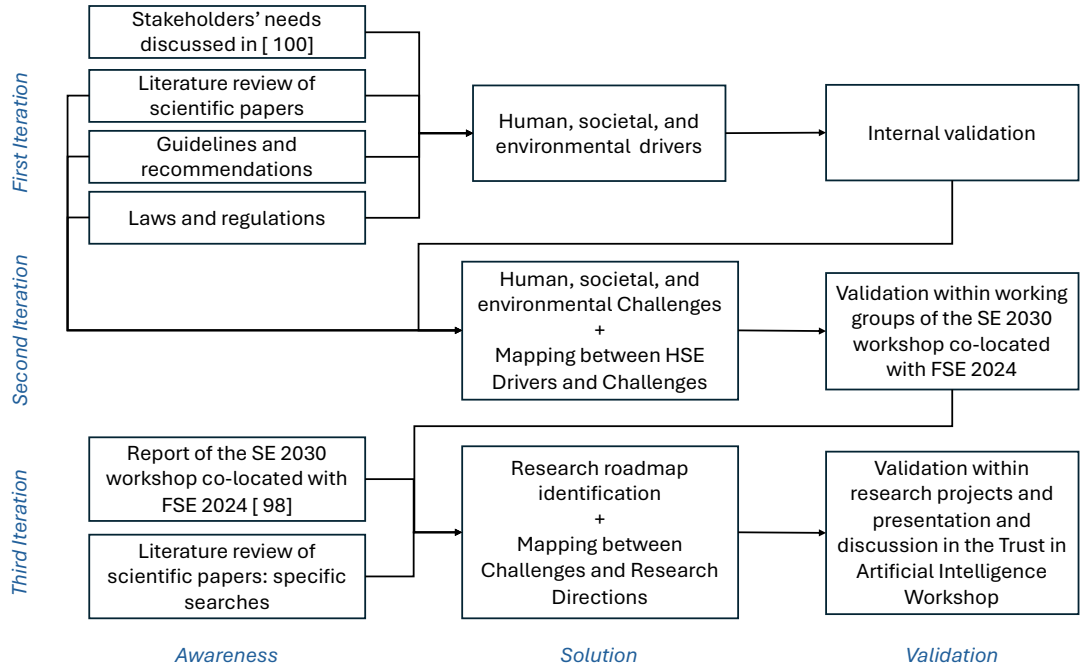

Fig. 1. Research methodology

- Needs of the stakeholders discussed in [100];
- A literature review and analyzed scientific papers in the field;
- Guidelines and recommendations, including guidelines from institutional bodies (e.g., UNESCO [117]), AI ethics frameworks (e.g., the Institute of Electrical and Electronics Engineers (IEEE) AI Ethics Framework [2], the European Union (EU) Ethics Guidelines for Trustworthy AI [58], the Organization for Economic Cooperation and Development (OECD) [93]), recommendations from different countries (e.g., the United States Government [119]);
- Regulations, including *GDPR* in 2018 and *AI Act* [77].

Then, we identified HSE drivers and validated them internally among the co-authors. Two co-authors independently identified the drivers and then converged to a final set of drivers with the supervision of a third co-author who also facilitated the discussion and integration of the sets of HSE drivers.

In the second iteration, informed by the validation of the first iteration, we extracted from the literature review, guidelines and recommendations, and laws and regulations HSE challenges, and we defined a mapping between HSE drivers and challenges. Concerning the validation phase, two co-authors participated in the two-day Software Engineering 2030 (SE2030) workshop co-located with FSE 2024, summarized in [98]. On July 15th and 16th, 2024, in Porto de Galinhas, Brazil, the event highlighted the software engineering research horizon, contributing ideas for the ACM TOSEM special issue. The report of the event [98] discusses seven new challenges to software engineering that emerged in the SE2030 workshop. They are very connected to the ambition of this paper, especially the software engineering by and for humans, and sustainable software engineering ones. However, also the other five challenges have been useful for identifying the research directions we formulated in this paper; they are AI for software engineering, automatic programming, cyber security, validation and verification, and quantum software engineering. The two-day workshop was organized into seven working groups by following a liberating structure[1]. Each working group met in various sessions, each with a duration of 30 minutes, followed by a plenary session where representatives of the groups reported findings and acquired knowledge to the other participants. Workshop participants were asked to rotate over time and participate in all groups. These working groups have been important in validating both the HSE challenges and mapping between HSE drivers and challenges.

The outcome of the validation of the second iteration and the SE2030 workshop have also been precious in identifying the open research questions and future research directions. In the third iteration, we also considered the summary and report of the SE2030 workshop [98], which was made available after the workshop and that perfectly complemented and confirmed the notes and acquired knowledge of the two co-authors that participated in the workshop. Based on this knowledge, we started identifying and framing the 14 research directions composing the roadmap. We also performed specific searches, one for each research direction, to identify relevant papers and to confirm or better discuss the research directions. Once the roadmap reached a good maturity level, we built a mapping between challenges and research directions.

Finally, we validated them in the context of three research projects in which the authors are involved. Halo ("HALO: etHical-aware AdjustabLe autOnomous systems") and Robochor ("RoboChor: Robot Choreography") are two Italian projects involving all the co-authors. In the case of Halo, we investigate how to "adjust" the autonomy of autonomous systems, i.e., how to shift autonomy from the system to humans based on ethical aspects. In the case of Robochor, we investigate how to coordinate a team of robots to achieve their mission. In both projects, we investigate means to specify not only the missions of robots but also the ethical preferences and properties of the involved humans. The other project, an internal project of the University of L'Aquila and Gran Sasso Science Institute (GSSI), is called EXOSOUL, and it aims at building a personalized software exoskeleton that enhances and protects humans by mediating their interactions with the digital world according to their own ethics of actions and privacy of data. The rationale behind the software exoskeleton is to disallow or adapt interactions that would result in unacceptable or morally wrong behaviors according to the ethics and privacy preferences of the human user. The project started over 10 years ago, involves all co-authors, and has weekly meetings to discuss related topics. The project is intrinsically multidisciplinary and involves philosophers, psychologists, sociologists and lawyers besides computer scientists and data scientists. This has been a perfect setting to test the research roadmap and provide an understanding beyond the purely technical aspects.

One of the co-authors also got invited to present the findings of this paper (talk titled "Engineering Digital Systems for Humanity") in a two-day workshop named "Trust in Artificial Intelligence" that aimed to explore the critical theme of trust in AI. The workshop was organized at the University of L'Aquila and brought together renowned experts to discuss AI's ethical, practical, and conceptual challenges. Three of the co-authors also participated in the event and the discussion sessions. This has been an opportunity to discuss our roadmap, even though more from the perspective of trust and AI.

**Threats to validity**

In this section, we discuss potential threats to construct, internal, and external validity of our study and how we mitigated them.

---

[1]https://www.liberatingstructures.com/12-2510-crowd-sourcing

*Construct Validity* – It refers to the extent to which our measurements and interpretations accurately reflect the concepts they intend to measure. In this study, construct validity could be influenced by the choice of sources of information to build HSE drivers, challenges, research roadmap, and mappings. To mitigate this threat, we carefully selected the sources of information by covering scientific papers, guidelines and recommendations, and laws and regulations.

*Internal Validity* – It concerns the extent to which the results can be attributed to the experimental manipulation rather than other factors. A potential threat is the interpretation of data acquired for the study and the consequent addition of biases in formulating the HSE drivers, challenges, and research roadmap. To address this, we always involved more than one co-author in every step, and we checked our findings with experts in the field, especially regarding guidelines and recommendations, laws and regulations.

*External Validity* – It concerns the generalizability of the findings beyond the study's specific context. Indeed, we cannot claim that the HSE drivers, challenges, and research directions are complete. We may have missed something since it was not covered in the sources of information we selected. It is also worth highlighting that the field is rapidly evolving, and there could be additional aspects to consider, e.g., due to new regulations or laws that may be proposed in the near future. Future work could enhance external validity, e.g., by considering documents available in the future or conducting an empirical study with domain experts.

## 3 BEYOND BUSINESS GOALS AND TECHNOLOGY DRIVERS: THE NEED FOR HUMAN, SOCIETAL, AND ENVIRONMENTAL DRIVERS

The various industrial revolutions have brought transformations in all societal systems. The evolutionary history of industry has now reached Industry 5.0. Its main objective is to prioritize human well-being within manufacturing systems, to foster prosperity for the sustainable development of all humanity [45]. Moving towards this new perspective is challenging, and the manufacturing paradigm needs time to adapt to the requirements of a novel society [71]. Society 5.0 aims at balancing economic advancement with the resolution of societal problems [62]. Sustainable computing is one of the terms used to describe the relationship between digitalization and sustainability. The Manifesto for Sustainability Design [17] highlights the societal and individual dimensions, among others. The social dimension focuses on societal communities and the factors that undermine trust in society, such as social equity, justice, employment, and democracy. The individual dimension pertains to the well-being of people, encompassing mental and physical health, education, self-respect, skills, and mobility [17]. The study in [53] identifies the sustainability needs of the industry concerning education and training from the perspective of software engineers. Industries, indeed, aim to enhance the sustainability of both software processes and products.

Emerging from the Vienna Manifesto on Digital Humanism [125] and the associated lecture series, the Digital Humanism initiative [70, 124, 126] heavily insists on human, societal and environmental values in digital systems. It deals with the complex relationships between people and machines in digital times. It acknowledges how digital transformation has opened the way to an extensive comprehension of informatics and IT from a mere engineering discipline to a worldwide endeavor touching every aspect of our lives. The initiative points to societal threats such as privacy violations and ethical concerns around digital transformation and advocates for Digital Humanism, emphasizing the need to shape technology in line with human values and universal rights to foster a better society and quality of life. At the same time, the initiative points out how the notion of sustainability needs to be reshaped and must evolve with digital technologies since if, on the one hand, the new digital technologies represent an opportunity, on the other hand, they can have a strong negative impact on the environment. From a variety of disciplines, including computer science, philosophy, education, and sociology, the thinkers involved in the initiative draw special attention to the fact that, as the digital transformation progresses, an environmentally sustainable digital society can become a reality only if we, as individuals and as a society, should and must make decisions taking democratic, humanistic, and environmental considerations into account.

Responsible computing in the digital world has also been thoroughly examined [65]. The author delved into the foundational principles of addressing ethical concerns in autonomous systems to mitigate the harm of digital society. Along the same line, Lu et al. identified a roadmap [76] as well as best practices [75] for responsible AI systems. Responsible AI is concerned with ensuring the responsible development and operation of AI systems. It has emerged as a significant challenge in the AI era, targeting both legal and ethical aspects that must be considered to achieve trustworthy AI systems. In [76], the authors defined a roadmap for software engineering methods on how to develop responsible AI systems, including requirements engineering, systems design, and operation. The roadmap also refers to ensuring the compliance of the development and use of AI systems for ethical regulations and responsibilities, and defining architectural styles for responsible AI-by-design.

The needs for architecting and engineering value-based ecosystems were also highlighted in [95]. In this work, the authors develop the concept of (dynamic) ecosystems as emerging from the mutual interaction of people, systems, and machines due to the continuous digitalization of the human world. These interactions can have different natures, such as collaborative, competitive, or malicious, and AI can further enhance them. In this context, humans could find themselves vulnerable in their engagements with the digital realm due to issues concerning societal values, such as fairness and privacy. To ensure a better digital society, the authors argue the need for engineering values-based digital ecosystems, equipped with built-in mechanisms to protect societal and human values.

Empowering the user with personalized software connectors that are able to reflect the user's moral values in the interactions with digital systems is the approach taken in the Exosoul project [9]. The approach has been applied to privacy profiles [66, 99] and to more general ethical profiles [6]. In line with this work, Boltz et al. [22], in their vision paper, emphasize the need for human empowerment within the framework of self-adaptive socio-technical systems, which demand mechanisms for balancing diverse needs, values, and ethics at the individual, community, and societal levels. Other approaches offer a different perspective for operationalizing values. Bennaceur et al. [18] promote the shifting of the definition and measurement of users' values from the early stages of software development to the runtime, supported by the software itself. Under the broad umbrella of dynamic ecosystems, we can categorize research about the human-machine teaming paradigm [33, 56]. While systems are still expected to operate autonomously, they are viewed as partners rather than tools in accomplishing mission objectives. To enable this transition, humans and machines must engage in closer interaction, fostering meaningful partnerships where decisions are collaboratively made [33].

Multiple countries worldwide are formulating and enacting legislation and policies concerning AI governance. In the USA, Biden enacted the "Executive Order on the Safe, Secure, and Trustworthy Development and Use of Artificial Intelligence" [120], whose purpose is that of promoting the responsible use of AI. Similarly, the Cyberspace Administration of China issued the final version of the "Interim Administrative Measures for Generative Artificial Intelligence Service" [31], establishing measures applying to the provision of generative AI services within China. The European Union has been the precursor to these legislations. Indeed, the European Parliament released the so-called *AI Act*, the first law to regulate the use of AI in Europe [77]. It provides a standardized framework for both the use and provision of AI systems within the European Union (EU). This framework delineates different requirements and obligations tailored to a risk-based approach and used to classify AI systems. The AI Act applies primarily to providers and deployers of AI systems and general-purpose AI models. Four categories of risk are identified, and legal interventions are tailored to these concrete levels of risk. AI systems presenting threats to people's safety, livelihoods, and rights are prohibited. Analogously to what happened with privacy and the European GDPR regulation, the EU has pioneered the AI regulation, and yet multiple countries worldwide are designing AI governance legislation and policies. As a result of the phenomenon known as the *Brussels effect* [23], compliance with EU laws extends even beyond its borders.

In the remainder of this section, we introduce some motivating examples to explain the concepts discussed in this paper (Section 3.1); we better explore the human role in digital systems (Section 3.2); and, finally, we present the human, societal, and environmental drivers (Section 3.3).

## 3.1 Motivating examples

Before describing the macro and technological challenges, we introduce motivating examples that will be exploited later as running examples mapping to the human roles in the interaction with digital systems, as detailed in Section 3.2.

**User-friendly Cobots** – This first example refers to the use of Cobots, i.e., collaborative robots intended for direct human-robot interaction within a shared space, or where humans and robots are in close proximity, in a production environment [42]. Our cobot is equipped with software enabling the robot to be programmed, i.e., the specification of the mission to be accomplished. The cobot provides multiple instruments to facilitate its programming by domain experts, who may not necessarily be experts in robotics or ICT. The first instrument is a textual and graphical domain-specific language (DSL) to enable a more "traditional" way of programming the robot. The second instrument allows programming the cobot using voice commands. Through the third instrument, the cobot learns by examples, either with the human showing by movements the desired behavior or by enabling learning from feedback on incorrect actions.

Programming is not completed before mission execution; rather, it is a continuous process that occurs as the cobot and humans collaborate in real time. In addition to safety mechanisms, such as safety-monitored stops and power or force limitations when required, the cobot is designed to be ethically aware, ensuring it adheres to ethical standards throughout its operations.

**Assistive robots** (inspired by [8]) – This example focuses on the use of assistive robots in nursing homes, where they support both caregivers and elderly residents. These robots are designed to engage in short, health-related conversations based on the user's condition, enabling them to assess the user's overall health status. Afterward, they can deliver necessary items, such as medication and water. To aid caregivers, the robot logs each interaction, allowing them to verify whether the patient has taken their medication. Assistive robots also conduct regular check-ups on residents with specific health conditions and alert nursing staff if any assistance is needed. By providing consistent, calm interactions, these robots can help reduce social isolation, distress, and confusion among older adults [105]. However, while they lack negative human traits like anger or frustration, they also cannot offer true compassion or empathy, presenting a potential limitation [105].

To ensure these robots enhance, rather than diminish, the quality of life for elderly residents, they must be designed with privacy and dignity in mind. This includes addressing concerns related to personal data, privacy rights, including the right to be let alone, and autonomy [88]. Additionally, assistive robots should monitor for signs of discomfort or reluctance in the residents, recognizing when the interaction may cause distress due to the absence of human care. In such cases, the robots must be capable of transferring control to human caregivers to fulfill needs requiring human empathy and compassion.

**AI for banking systems** – The increasing use of digitalization and AI in financial services, such as banking and insurance, is transforming the industry. As highlighted by Moody in a recent global study on AI adoption in risk management and compliance [86], these technologies are now integral to the operations of financial institutions. Today, customers can effortlessly log into mobile banking apps, check their account balances, and carry out various tasks while AI algorithms analyze their transactions, spending habits, and financial goals. These AI-driven systems are employed to enhance efficiency, improve transaction speed, and reduce operational costs [86]. Based on the data gathered, the AI can offer personalized financial advice, ranging from budgeting tips to investment opportunities, tailoring the experience to each user's unique financial profile. On one hand, AI enriches the banking experience by delivering customized recommendations and robust security measures to ensure customer satisfaction. On the other hand, these systems are vulnerable to fairness concerns, such as biases in data or AI models. For example, AI algorithms used in loan approval processes may inadvertently discriminate against customers based on nationality, gender, or other factors, often without explicit awareness of such biases.

To mitigate these risks, AI-enabled systems must be carefully designed to prevent unfair behavior, protect privacy, and eliminate bias [52, 80], ensuring ethical and equitable financial services for all users.

## 3.2 Human role in digital systems

Modern systems in a digital society are crucial to enhancing and streamlining daily human activities. By leveraging AI technologies as key enablers, these systems often operate autonomously and adapt to individual needs and preferences. They enable both *explicit* and *implicit* interactions, providing a range of automated services and processes that improve efficiency. As a result, they play a pivotal role in shaping our interactions with the (digital) world and each other, transforming how we live and work. Additionally, due to their widespread diffusion, these systems significantly influence societal, economic, and political dynamics. The literature on human-system interactions explores various roles that humans can play in their interactions with systems, typically categorized as proactive, reactive, or passive.

When playing a **proactive role**, humans *take the initiative, proactively,* to engage with systems to achieve specific goals or to influence the system's behavior, rather than just responding to it. This often involves anticipating needs and making adjustments or inputs before the system requires intervention. Literature in this area often focuses on how systems can be designed to support proactive behaviors and interactions [13, 19, 94]. The user-friendly Cobots previously mentioned in Section 3.1 are also examples of systems where the user proactively engages in interacting with the cobots by showing through movements or by means of feedback the desired and/or undesired behavior.

A **reactive role** refers to humans engaging with systems in real time to perform tasks, make decisions, or provide input. This involves direct, hands-on interaction with the system, such as using software applications, controlling devices, or inputting data. While users are actively involved in this role, their engagement is typically *reactive*, responding to events or system prompts, rather than taking proactive actions [69]. The assistive robots previously described in Section 3.1 are examples of systems where the user actively participates in interacting with the robots by responding to requests or reacting to events.

In a **passive role**, humans interact with systems *indirectly*, often providing little to no direct input. These systems may collect data or respond to user behaviors without requiring active engagement. Research on passive interaction examines how systems can autonomously monitor and adapt to user behaviors, as seen in examples like ambient intelligence [110]. However, it is important to consider the potential privacy concerns that arise from systems operating with minimal user intervention, as they may collect and process personal data without explicit user control. The AI for banking systems previously described in Section 3.1 are clear examples of systems where the user passively participates in the interaction with the systems, and these interactions are greatly reduced.

Although the literature emphasizes the importance of designing systems that accommodate and enhance various interaction roles to improve usability, efficiency, and user satisfaction, the significance of human-system interactions extends beyond system usability [111]. These interactions can have profound ethical implications, often influencing users in ways they may not even be aware of. Systems that adapt to different interaction styles may subtly shape user behavior, decisions, or perceptions, raising ethical concerns about autonomy, privacy, and consent, even when users believe they are passively engaged. Additionally, in modern digital systems (e.g., socio-technical systems), everyone is a user. These systems exist and operate within society and evolve alongside us. As a result, the concept of a "user" extends beyond the traditional idea of the "end user", encompassing all of us in a broader, more integrated sense.

## 3.3 Human, societal, and environmental drivers

Overall, the analyzed works delineate the need to consider human, societal, and environmental values while engineering systems. As a matter of fact, the traditional engineering of systems is typically taking into account business goals [16, 30, 34, 35, 54], intended as the higher purpose and added value of a company, and technology drivers [1]. Business goals have an important role in identifying architectural requirements, i.e., requirements that profoundly impact the architecture of a software system [16]. The importance of aligning business goals with the architecture and the system under development is widely highlighted in the literature [26, 30, 34, 35]; as underlined in [30], there is almost the "near-universal acceptance of the importance of business/IT alignment". Recently, this problem has also been investigated by Scaled Agile [102]. Business goals have the objective to (i) maintain growth and continuity of the business, (ii) meet the company's financial objectives, (iii) meet personal objectives, (iv) meet the responsibility to employees, (v) meet the responsibility to society, (vi) meet the responsibility to countries, (vii) meet the responsibility to shareholders, (viii) manage market position, (ix) improve business processes, and (x) manage quality and reputation of products [35]. Examples of business goals in the automotive domain are, for instance, product evolution after-sale and increased OEM control over unknown/concerns [26]. Technology drivers are forefront technologies or processes that lead to the development of next-generation equipment together with development and manufacturing processes and techniques [1]. Examples of technology drivers in the automotive domain are, for instance, autonomy intelligence, software as a service, and software over-the-air updates [26].

The work in [35] highlights several critical factors that must be considered when architecting software systems. These include changes in various environments such as the social, legal and regulatory, competitive, technological, and customer environments. The engineering of digital systems for humanity must take into account HSE factors. These drivers play a critical role in shaping digital systems that are not only technologically sound but also socially responsible. To describe the main HSE drivers, we refer to the stakeholders' needs discussed in [100]. The authors retrieved them via a literature review including guidelines from institutional bodies (e.g., UNESCO [117]), AI ethics frameworks (e.g., the Institute of Electrical and Electronics Engineers (IEEE) AI Ethics Framework [2], the European Union (EU) Ethics Guidelines for Trustworthy AI [58], the Organization for Economic Cooperation and Development (OECD) [93]), recommendations from different countries (e.g., the United States Government [119]), and recent literature. The elicited drivers include:

- **D1:** *Societal and environmental well-being* – Systems should be engineered to respect, enhance, and benefit society as a whole, including individuals and the environment [2, 58, 93, 117].
- **D2:** *Accountability* – Systems should be engineered in a manner that enables clear responsibility for their behavior and outcomes to be established [2, 58, 93, 117].

- **D3:** *Privacy and data governance* – Data should be managed in such a way that accuracy, completeness, consistency, validity, timeliness, and uniqueness are ensured in real-world data [75].
- **D4:** *Human agency and oversight, human autonomy* – Systems should be engineered in a manner that ensures humans retain control and decision-making authority [58, 72, 117, 119].
- **D5:** *Transparency, Explicability, and Explainability* – Systems should be designed in such a way that they facilitate the ability to explain both the technical processes of an AI-enabled system and the related human decisions (e.g., application areas of a system). Explainability should also ensure that humans have the right to know they are interacting with a system [2, 12, 58, 72, 75, 93, 117, 119].
- **D6:** *Diversity, non-discrimination, and fairness* – Systems should be designed in such a way that they guarantee the avoidance of unfair bias in datasets used by AI systems, as well as in AI models. Accessibility should also be prioritized to ensure user-centricity and ease of use for all individuals and any group of people, regardless of age, gender, or abilities [12, 58, 72, 75, 93, 117, 119].

It is worth noting that HSE drivers could, and likely will, become significant business drivers in the future. We can observe a similar phenomenon in green values: environmental values have become business values, and companies are increasingly using them in their advertisement and marketing strategies. An analogous situation is likely occurring with sustainability and its various dimensions. As discussed in [53], organizations aim to enhance the sustainability of software processes and products but face challenges such as balancing short-term financial profitability with long-term sustainability goals and dealing with an unclear understanding of sustainability concepts from a software engineering perspective. We already observe ethics in business as a means to help companies build trust with their customers, employees, investors, etc. When businesses are perceived as ethical, they enhance their reputation by fostering stronger customer loyalty, boosting employee morale and productivity, and reducing the risk of legal and regulatory issues. What we foresee now is something slightly different and more focused on technology. Customers will soon become more willing to buy products based on AI when the producers provide guarantees of respecting HSE values. This is also what Bart Willemsen, research vice president at Gartner, anticipated in 2021 concerning privacy: "Even where regulations do not yet exist, customers are actively choosing to engage with organizations that respect their privacy" [108].

## 4 MACRO AND TECHNOLOGICAL CHALLENGES

Based on the context presented and the examples discussed in Section 3, we identify four major technological challenges together with their subchallenges, as schematized in Figure 2. Figure 3, instead, shows the mapping between the HSE drivers and the identified challenges.

### 4.1 Continuous systems programming (Proactive human role)

As already introduced, this challenge relates to humans being proactive rather than reacting to events that happen, i.e., taking the initiative to ask robots to perform specific tasks. We see this activity as a sort of programming that is more than just interaction, which will be the focus of the next challenge. Humans should then be able to program systems continuously, from design time to runtime, by altering, reprogramming, or changing their behavior. First, it is crucial for humans to be able to program the system by specifying its behavior in specific situations. This programming phase is distinct from the system's programming phase during production, underscoring the unique role of developers. It is a post-production programming phase that is carried out by users in the context of use. It can go from the settings of some parameters to a richer specification of the system behavior,

```
┌─────────────────────────────────────┐  ┌──────────────────────────────────────────────┐
│ CH1: Continuous system programming  │  │ • CH1.1 - Ease of use and accessibility      │
│   • Human role: Proactive role      │──│ • CH1.2 - Continuous monitoring and assessment│
│                                     │  │ • CH1.3 - Continuous compliance              │
└─────────────────────────────────────┘  └──────────────────────────────────────────────┘

┌─────────────────────────────────────┐  ┌──────────────────────────────────────────────┐
│ CH2: Human-systems interaction      │  │ • CH2.1 - Elicitation of values              │
│   • Human role: Reactive role       │──│ • CH2.2 - Redistribution of autonomy         │
└─────────────────────────────────────┘  └──────────────────────────────────────────────┘

┌─────────────────────────────────────┐  ┌──────────────────────────────────────────────┐
│ CH3: Digital systems impact on humans│ │ • CH3.1 - New quality standard               │
│   • Human role: Passive role        │──│ • CH3.2 - Multidisciplinarity and new        │
│                                     │  │   development methods                         │
└─────────────────────────────────────┘  └──────────────────────────────────────────────┘

┌─────────────────────────────────────┐  ┌──────────────────────────────────────────────┐
│ CH4: Trust and trustworthiness      │  │ • CH4.1 - Methods for enabling trust         │
│   • Human role: Any                 │──│ • CH4.2 - Measuring quality                  │
│     (Duality of rigorous design     │  │ • CH4.3 - Testing of autonomous systems      │
│      and subjective belief)         │  │                                              │
└─────────────────────────────────────┘  └──────────────────────────────────────────────┘
```

Fig. 2. Main Challenges for Engineering Digital Systems for Humanity

what should be avoided, and the specification of quality and ethical aspects [9, 37, 81, 83, 112]. More often than not, the persons that should accomplish this programming phase are not developers and, therefore, the programming language should be accessible and easy to use by stakeholders that are experts of the application domain but not experts in ICT. In the case of the user-friendly cobots motivating example, the humans that should program the cobots are experts in the domain of manufacturing, e.g., in automotive manufacturing, but they are not necessarily experts in robotics. The language should be user-friendly and intuitive but at the same time unambiguous and rigorous, and the terminology should be the same as the domain. In the case of the motivating example of assistive healthcare robots, the language should enable the patient, or operators supporting the patient in this activity, to specify preferences like consensus about the treatment, risk management assessment, balance for benefit, or moral attitude. In the literature, various approaches aim to make mission specifications accessible to users without ICT expertise, empowering them to accurately and effectively define the tasks that robots should perform. Patterns for mission specification have been identified by surveying the state of the art and formulating catalogs of specification patterns for mobile robots [84, 85]. Catalogs come with a structured English grammar enabling the specification of the mission in English. This user-friendly specification of the mission is then automatically transformed into a temporal logic specification, thus enabling the use of various tools for, e.g., synthesizing planners or controllers or enabling verification. Other researchers have developed domain-specific languages (DSLs) to provide user-friendly methods for programming robots [38, 39]. Other approaches propose gesture-based programming for robots within the context of human-robot collaboration in shared work environments [92], and cobot programming for collaborative industrial tasks [42]. In [92], the authors present a ROS-based software called MEGURU, where a user-friendly gesture language is developed using gestures that are easy for users to perform and convey a clear, intuitive command. A survey on intuitive cobot programming is reported in [42].

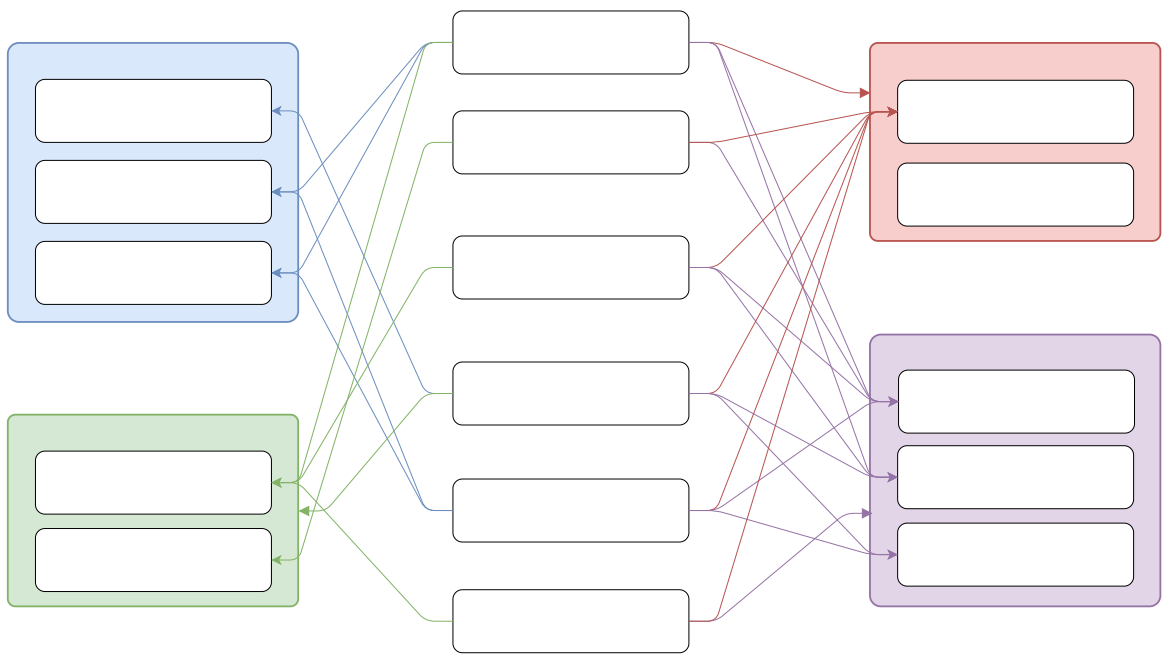

Fig. 3. Mapping between HSE Drivers and Challenges

Besides ease of use and accessibility, it also becomes important to put in place mechanisms for continuous monitoring and assessment to enable stakeholders to build confidence that the system behaves correctly and satisfies the quality aspects that stakeholders identify as important. Continuous monitoring and assessment can also be exploited to build approaches to guarantee compliance with standards or regulations continuously so that each new version of the software (deployed when the system is in the field) has the same safety, security, and ethical guarantees as previous versions.

To facilitate the transition towards meaningful partnerships between humans and systems, where decisions are collaboratively made also through continuous programming, multiple challenges must be considered.

> **Challenge:**
> - **CH1.1:** *Easy of use and accessibility* – Engineering for humanity, especially taking into account the human values, requires a proper balance between simplicity/flexibility/accessibility and rigorousness/unambiguity/accuracy to enable humans, non-expert in ICT, to confidently and correctly program systems in a continuous fashion. There is a need for domain-specific languages that are simple and intuitive yet rigorous enough to guide users in accurately specifying their needs. Additionally, other programming methods should be explored, such as imitation learning for example-based programming, as well as leveraging large language models to facilitate programming through voice and speech.
> - **CH1.2:** *Continuous monitoring and assessment* – The evolution of digital systems after production and their adaptation to operational environments necessitate new development methodologies that encompass activities beyond production, including maintenance, evolution, and continuous monitoring and assessment. This is also made clear by regulations regarding AI governance, such as the AI Act, which include provisions for post-market monitoring of both the use and provision of AI systems based on risk assessments. This challenge is relevant to all the values discussed in this paper, i.e., human, societal, and environmental values.
> - **CH1.3:** *Continuous compliance* – When continuous systems programming is implemented, it extends beyond internal activities in the companies, such as continuous monitoring and assessment. Continuous compliance with safety and security regulations, as well as with the AI Act or other AI regulations, must be ensured to both protect humans participating in a proactive role and prevent malicious behaviors. This shift poses challenges to traditional certification and compliance approaches. This challenge is relevant for human, societal, and environmental values because of the potential and broad impact of non-compliance.
>
> <div style="text-align:right">**CH1: Continuous systems programming**</div>

## 4.2 Humans-systems interaction (Reactive human role)

This challenge concerns humans having a reactive role while interacting with digital systems. One important aspect to consider is that humans and systems need to share the same environment, not necessarily physical, but also digital or hybrid. This is, for instance, the case of Assistive Robots. This implies that other ethical values and human dignity come into play besides privacy concerns. *Digital ethics* is the branch of ethics concerning the study and evaluation of moral problems related to data and information, algorithms, and corresponding practices and infrastructures, aiming to formulate and support morally good solutions [49]. Digital ethics is made by *hard ethics* and *soft ethics*, which are sometimes intertwined inextricably. System producers may already adhere to the hard ethics rules defined by legislation, which are considered collectively accepted values. Soft ethics define personal preferences, and it is the responsibility of soft ethics to shape the users' interaction with the digital world [9, 65, 82]. The subjectivity in the ethics specification is also confirmed by the results of the moral machine experiment, which highlights how demographic and cultural traits affect moral preferences [11]. These considerations highlight the need for engineering solutions for flexible, customizable, and privacy-preserving interactions between humans and systems. Humans should be able to express and negotiate their preferences, while systems should dynamically adhere to the diverse soft ethics of the users they interact with [81, 83]. This calls also for emphatic systems, i.e., systems able to understand, interpret, and respond to human emotions [109]. In other words, there is the need for instruments to guarantee that the behavior of the system, e.g., robot(s), will be compliant with the specified ethical preferences (as discussed in Section 4.1). The use of synthesis techniques can then help to automatically generate the correct-by-construction logic needed for coordinating the robots and their interactions with humans, as well as the environment, in a way that the specified mission is accomplished in the correct and morally good manner [9, 65]. Synthesis techniques can take as input the mission specification, including the ethical preferences, and can automatically generate a controller able to mediate the interactions between robots and humans so as to guarantee the accomplishment of the mission, when possible, while protecting and preserving ethical preferences. Also, the controller should be able to understand the situations in which a redistribution of control from robots to humans is needed to not violate human ethics [46]. Adjustable autonomy is the means to redistribute the operational control among different parts of the system, as well as humans [87]. Alfieri et al. [5] investigated the literature with the aim of clarifying the distinction between human replacement and human augmentation, in the context of autonomous intelligent systems. They observed a prevailing negative perception regarding human replacement, whereas there is a generally favorable attitude towards enhancing them.

To facilitate the transition towards ethical-aware interactions between humans and systems, where decisions are made by also trying to protect and preserve ethical preferences, multiple challenges must be considered.

> **Challenge:**
> - **CH2.1:** *Elicitation of values* – Eliciting human, societal, and environmental values is an activity that cannot be fully anticipated during the design phase, as accurately profiling individuals based on their moral preferences is inherently challenging [65]. Moreover, due to the evolving nature of humans and their awareness, the elicitation and specification of ethics should be done in a continuous fashion. AI is a promising instrument for inferring not only ethical preferences but also emotions in a continuous way. However, this instrument could be itself too risky for human beings. For example, as far as emotions are concerned, the recent AI Act of the European community *"[...] prohibits placing on the market, putting into service [...], or use of AI systems to infer emotions of a natural person in the areas of workplace and education institutions except in cases where the use of the AI system is intended for medical or safety reasons."* [77].
> - **CH2.2:** *Redistribution of autonomy* – Developing systems capable of measuring and automatically redistributing autonomy and control presents significant challenges. This complexity arises from the need to balance individual preferences and well-being with collective needs, ensuring that such systems can adapt to diverse contexts and maintain fairness while promoting effective decision-making. This challenge is relevant for the human and societal values, intended as the individual preferences vs. the collective needs. However, the collective needs can also include the environmental values.
>
> <div align="right">**CH2: Humans-systems interaction**</div>

## 4.3 Digital systems impact on humans (Passive human role)

This challenge focuses on the impact digital systems can have on humans merely playing a passive role, i.e., the interaction with systems does not involve active participation. These digital systems must be engineered to guarantee fair and ethical decision-making processes and comply with ethical standards and regulations. This is a crucial aspect since their users might be vulnerable and unaware that their values could be at risk, like in the AI for banking system example. One aspect that requires attention is removing bias from systems (e.g., processing, design, algorithm itself) and from the data used to train AI-based systems [29, 61, 80]. This is important because bias could lead to unfair behaviors of the systems with potential discrimination for individuals, groups, or subgroups of people [36]. Enhancing explainability of AI and systems' behavior is widely investigated to provide greater transparency [41, 52, 78, 114]. However, in the literature, we can find various terms that often overlap each other, such as explainability, transparency, interpretability, and understandability. Software engineering can help to clarify these concepts [121], similarly to what is done for more traditional quality attributes [107].

To support the transition towards explainable and transparent systems while also increasing users' awareness, diverse challenges must be considered.

> **Challenge:**
> - **CH3.1:** *New quality standard* – This macro challenge encompasses various aspects, including explainability, transparency, interpretability, and understandability. Greater conceptual clarity is needed to prevent misunderstandings in AI systems. Additionally, the community would benefit from a new or revised quality standard that incorporates these emerging qualities, which are becoming essential as AI systems increase in complexity and play a larger role in critical decision-making processes. This would not only enhance accountability but also improve the overall usability and reliability of AI systems. Since this challenge focuses on the new quality standard, it is relevant for all (human, societal, and environmental) values.
> - **CH3.2:** *Multidisciplinarity and new development methods* – Regulations such as the AI Act introduce new requirements for high-risk and limited-risk AI-based systems, extending beyond purely technical aspects. For instance, the transparency requirement mandates informing users about their rights and the fact that they are interacting with an AI-based system. As AI continues to integrate into everyday life, there will be a growing need for comprehensive solutions and new development methods that address not only technical challenges but also ethical, legal, and social dimensions. To achieve this, a multidisciplinary approach is essential, bringing together expertise from fields such as law, ethics, social sciences, and psychology alongside engineering and computer science. This collaboration will help create digital systems that are not only technologically advanced but also aligned with societal values, human rights, and the well-being of individuals. Such methods will be crucial for designing and developing systems that can be trusted, understood, and responsibly deployed in ways that benefit humanity as a whole. New development methods should address all the relevant values, encompassing human, societal, and environmental aspects.
>
> <div align="right">**CH3: Digital systems impact on humans**</div>

## 4.4 Trust & trustworthiness

Trust and Trustworthiness are cross-cutting challenges that have already been touched on in the previous challenges. In fact, the importance of *trust* and *trustworthiness* within these systems is increasingly compelling, as these are central aspects, especially with respect to the humans' interplay with digital systems. The new quality aspects that are increasingly important for digital systems have already been discussed in the previous challenges, like the need for new quality standards highlighted by CH3.1. In this challenge, we focus on instruments to achieve trust and trustworthiness.

Before detailing these challenges, we would like to highlight the dichotomy between the terms trust and trustworthiness, which are frequently treated as synonyms despite their distinct meanings. As discussed in [75], trust concerns the subjective acceptability of systems from the humans' perspective, whereas trustworthiness concerns the designing of systems in order to behave safely and guarantee security or quality aspects.

**Trust.** As a society, our dependence on intelligent machines is growing. Over the past few years, research has focused on addressing the question of how an external observer perceives systems. In this direction, enhancing *explainability* of AI and systems' behavior is widely investigated to provide greater transparency. In addition, modern digital systems are increasingly making autonomous decisions affecting our lives, such as hiring employees, granting loans, and so on. In such contexts, biases in software and AI significantly impact fairness. Consequently, much research is conducted to mitigate biases [29, 80] as well as to make AI explainable [41, 52]. Together with explainability, transparency is another quality of systems that promote trust [121]. *Transparency* is also highlighted in the AI Act [77], which provides various facets of transparency: (i) AI systems should be developed and used in a way that allows appropriate traceability and explainability, (ii) humans should be made aware that they communicate or interact with an AI system, (iii) deployers should be made aware by providers of the capabilities and limitations of the AI system, and (iv) affected persons should be made aware about their rights (such as the rights to free movement, non-discrimination, protection of private life and personal data).

In their recent work, Alfieri et al. [5] investigated the literature with the aim of clarifying the distinction between *human replacement* and *human augmentation*, in the context of autonomous intelligent systems, to which humans increasingly delegate important tasks and decisions. They observed a prevailing negative perception regarding human replacement, whereas there is a generally favorable attitude towards enhancing them. Concerns exist about a future dominated by machines, even if people appear to discern between beneficial and detrimental replacement scenarios [5]. For instance, the case of human caregivers substituted by care robots, such as the assistive robots scenario described in Section 3.1, is perceived as a negative scenario of replacement. The aforementioned adjustable autonomy, as a means to redistribute the operational control among systems and humans [87], is also motivated by the need for ways to deal with the absence of trust and confidence of human users.

**Trustworthiness.** Autonomous systems are considered trustworthy when they behave safely, and their design, engineering, and operation guarantee security or quality aspects [89]. This can depend on many factors, such as accountability, robustness, adaptation ability in dynamic and uncertain environments, security against attacks, and so on. Common to the various known techniques for trustworthiness demonstration, from synthesis to verification and testing, there is the need to formulate specifications for validating them [3]. However, demonstrating the trustworthiness of autonomous and/or AI-powered systems poses new challenges. For instance, testing AI systems is challenging due to their complex nature, which makes conventional testing approaches insufficient or impractical for these systems [91]. Also, testing approaches should move in the field, and they should be able to self-adapt to adjust (e.g., in terms of test cases, testing strategy, oracle) to the evolutions and adaptations of the system under test [106].

Multiple challenges arise when promoting trust and trustworthiness in digital systems, spanning from trust-building strategies to methods for measuring qualities to testing approaches.

> **Challenge:**
> - **CH4.1:** *Methods for enabling trust* – There is a growing need for methods that help humans build confident trust in digital systems. These methods can rely on metrics to measure quality attributes, ensuring system reliability and performance. However, trust-building extends beyond technical measures; it also requires intuitive and human-centered communication strategies. This is, however, difficult. For instance, an attempt in this direction is represented by the privacy labels introduced in December 2020 by Apple in its app store. Privacy labels are published on each mobile app's page and provide an easy and recognizable overview of data collection practices employed by the app, displaying what data is collected, how it is used, and for what purposes it is needed. This trust-enabling initiative towards transparency and understandability was also meant to discourage developers from collecting unnecessary data. However, as found by the large-scale study in [104], developers were not sensibly discouraged, and the majority of sensitive information was still collected in an anonymized form, a wide range of which for tracking purposes. In April 2021, Apple introduced further privacy-oriented changes to the iOS platform. After this release, app developers are required to explicitly request permission to track the user beyond the app in use. The introduction of these additional run-time checks, which return control back to end users, finally resulted in a statistically significant decrease in the number of apps that collect data for tracking purposes [104]. More straightforward positive examples of building trust are that a car could "smile" via a display to signal pedestrians that it is safe to cross, or a robot could exhibit empathetic behaviors, even if such behaviors are not explicitly defined in its mission. By integrating these human-like interactions, digital systems can foster trust by aligning with human expectations and emotional responses. Since this challenge focuses on methods for enabling trust, it is relevant to both human and societal values, as trust also has a societal dimension.
> - **CH4.2:** *Measuring quality* – The new quality dimensions discussed in Section 4.3 are often more challenging to measure than traditional attributes due to their subjective nature, e.g., soft ethics, and their dynamic characteristics, e.g., the evolving understanding of users over time. There is also a need for tools that simplify the communication of desired quality levels to users. One potential approach could be to introduce a system of discretization into categories or value ranks, similar to how energy efficiency classes are assigned to devices. These tools would allow users to confidently and accurately compare services based on these quality measurements, ensuring they make informed choices aligned with their preferences and values. Since this challenge focuses on means to measure quality, it is relevant for all (human, societal, and environmental) values.
> - **CH4.3:** *Testing of autonomous systems* – Adequate testing approaches are essential for autonomous systems, requiring equally robust specifications that can fully capture their complex nature. These specifications must address the multifaceted aspects of autonomous systems, including unpredictability, adaptability, and the interaction between the system and its environment. Effective testing methods should ensure that autonomous systems perform reliably, ethically, and safely across diverse scenarios. Since this challenge focuses on testing autonomous systems, it is relevant for all (human, societal, and environmental) values.
>
> **CH4: Trust & trustworthiness**

## 5 RESEARCH ROADMAP OF DIGITAL SYSTEMS FOR HUMANITY

In this section, building on the HSE drivers presented in Section 3 and the macro and technological challenges presented in Section 4, we identify a research roadmap of digital systems for humanity. The research roadmap is concretized in a number of research directions that are organized into four groups, namely, development process (Section 5.1), requirements engineering (Section 5.2), software architecture and design (Section 5.3), and verification and validation (Section 5.4). Figure 4 provides an overview of the challenges and the identified research directions while offering a mapping among them.

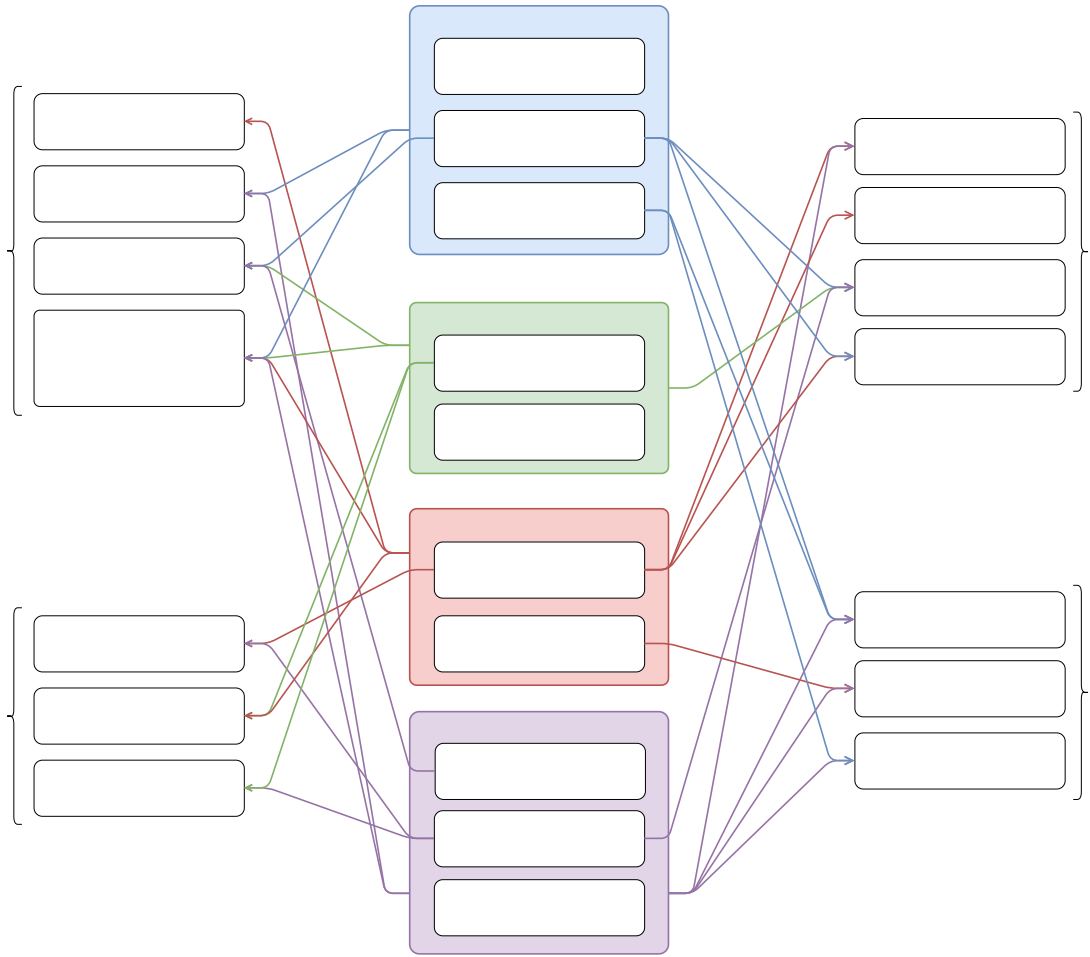

Fig. 4. Mapping between HSE Challenges and Research Directions

## 5.1 Development process

This subsection identifies four research directions that pave the path towards new software development processes for building digital systems for humanity. The first research roadmap, RD1.1, focuses on the need for cross-disciplinary collaborations and unavoidable multidisciplinary in order to cover the various competencies needed. RD1.2 emphasizes users taking a proactive role, as discussed in Section 4.1, by acting as programmers and configurators of autonomous or semi-autonomous systems in a continuous fashion. RD1.3 takes into account the subjective dimension of HSE values and investigates how systems should be built in order to adapt and configure to this subjectivity at runtime. Finally, RD1.4 investigates and poses questions concerning the role universities and, in general, research in software engineering should have in the era of automation, low-code platforms, and generative AI.

- **RD1.1:** *Cross-disciplinary collaboration and unavoidable multidisciplinary* – The introduction of ML components in system development has already challenged the reuse of traditional, non-ML software development processes [90]. In fact, ML processes require different practices in requirement engineering, design, testing/quality, process, and management [90]. In addition, ML system architectures typically include data gathering, data cleansing, feature engineering, modeling, execution, and deployment [90].

We now envisage the participation of stakeholders who are experts in HSE factors, such as philosophers, lawyers, and sociologists, in the development process. This introduces new challenges, particularly related to differences in culture, experience, instruments, and working styles. It is also crucial to determine how these experts should be integrated—whether as consultants on demand or as continuous participants in cross-functional teams throughout the development phases, following agile development practices. This highlights the need to investigate new tools that enable cross-disciplinary collaboration, providing software development solutions that support seamless communication and cooperation between experts from diverse backgrounds throughout system engineering.

- **RD1.2:** *Seamless DevExe and users as programmers* – As discussed in Section 4.1, continuous systems programming relates to the post-delivery ability of end-user developers (i.e., humans who are not professional software developers) to program the system by specifying its behavior for specific usages in specific situations. Humans are proactively enabled to program the system, from design time to runtime, continuously. The work in [13] classifies end-user development and programming techniques, including rule-based technique, programming by demonstration/example, wizard-based, natural language, etc. The work in [94] provides a historical evolution of end-user development and discusses challenges and possible research direction in the field, which include (i) providing support for managing complex, real-life personalization, (ii) offering the most appropriate abstraction level to end-user developers, (iii) providing users with effective means to understand and validate the correctness of the specified behavior, and so on.

    We foresee new programming paradigms and supporting tools for human-system interactions could be advantageous for tailoring HSE individual preferences. Since end-user programming is likely to contain errors due to the lack of expertise, novel programming systems are needed to assist users in identifying errors [4] and developing correct, yet more optimal, programs capable of properly accounting for the specified HSE values. Seamless *DevExe* processes are then needed, i.e., innovative development processes that blur the boundaries between development and execution while going beyond existing methods and techniques to empower users in modifying and creating digital artifacts, program things, and robots. In order to concretely support engineering for humanity, DevExe processes must enable a seamless transition that extends to V&V, maintenance, continuous monitoring, assessment, and compliance. Also, evolution should integrally become an indistinguishable phase that permits seamlessly re-entering the process from any phase during system execution.

- **RD1.3:** *Building systems that adapt and configure to human subjectivity at runtime* – Human values, along with social and environmental values, have a subjective dimension, as also highlighted in [95]. As described in Section 4.2, ethics has a clear subjective aspect. Consequently, digital systems designed for humanity should be capable of adapting their interactions to accommodate human subjectivity –such as avoiding or modifying behaviors based on the individuals interacting with the system at a given moment– and configuring themselves to meet the needs and profiles of specific users throughout the system's lifecycle. For example, rented cars should adjust their behavior according to the preferences of the passengers. This also calls for runtime negotiation with various and potentially conflicting human profiles, as discussed in RD2.3. The subjectivity of systems' behavior is a source of variability that must be resolved at runtime.

    To better explain this research direction, we focus on the service robotic domain. The work in [50] identifies the main drivers of variability as the environment, hardware, and mission. The environment also includes human interaction, but it mainly consists of innovative ways to guarantee safety and adherence to safety standards when robots operate in unpredictable

and uncontrolled environments. When considering human values, the interaction between robots and humans must also account for the variability in robot behavior, such as adapting to avoid causing distress, as discussed in the assistive robot motivating example.

There is a need to understand the precise limitations of existing methods and tools for runtime adaptation and configuration when addressing subjective human values and to develop new approaches to overcome these limitations.

- **RD1.4:** *The evolving role of universities and software engineering research in the era of automation, low-code platforms, and generative AI* – Software engineering research is certainly transforming in response to advancements in automation, low-code platforms, and generative AI. These new technologies are being integrated into different research areas, such as automated code generation, software testing, requirements elicitations, and system design, by disrupting traditional processes. This highlights the need to investigate how these disrupting technologies are influencing the future of the field. The scenario is even more complex, considering that AI technologies deployed in the industry tend to be more advanced than those developed in university laboratories [74]. As a result, a few companies are making faster progress and achieving more substantial breakthroughs than academic institutions. This concentration of power among a small number of actors amplifies their influence and raises concerns about potential risks to human, societal, and environmental well-being. Recently, prominent AI scientists have expressed concerns that the loss of human control or the malicious use of AI systems could result in catastrophic consequences for humanity as a whole [79]. In this evolving context, it is essential to reconsider the role of universities and research, to better align with the evolving landscape and prevent being left behind.

  Reflecting on the current circumstances, two fundamental roles for universities can be envisioned. A broader role for universities involves contributing to AI governance [21] by identifying ethical issues arising from the widespread use of digital systems [65], highlighting challenges overlooked by major industry players [74], and safeguarding human values through institutional activities. The study in [51] clearly reveals that people trust more national universities, research institutions, and defense organizations than governments and commercial organizations. A concrete and specific example of researchers involved in AI governance is given by individual academics and universities serving as external experts in the development of government policies and regulations, as occurred in the case of the AI Act [77].

  A second and more specific role involves directing research efforts towards the development of approaches and frameworks aiming to protect humans [72, 75, 123]. So far, approaches proposed in this direction deal with the encoding of user values [37], safeguarding humans in their interactions with digital systems [9], defining best practices for creating trustworthy AI systems [75]. Furthermore, there is a need to explore how universities, industry, and governments can collaborate on AI technologies in a socially responsible way [74].

## 5.2 Requirements Engineering

This subsection identifies three research directions concerning the engineering of requirements of digital systems built for humanity. The first research direction, RD2.1, suggests reconsidering quality to include new quality dimensions. RD2.2 focuses on the specification of HSE requirements, which is needed in order to enable, e.g., formulation, verification, and enforcement of such requirements. RD2.3 starts from the classical tradeoff of functional requirements, like performance vs security, and observes how the tradeoff concept should evolve towards runtime negotiation of HSE requirements.

- **RD2.1:** *Reconsidering quality to include new quality dimensions* – The introduction of legislation and policies concerning AI governance, which is spreading in different countries, implies the need for modern digital systems to be compliant with these legislations. Nevertheless, software or product quality and existing quality standards, for instance, the ISO/IEC 25010:2023 [67], go beyond compliance. They determine which quality characteristics will be taken into account when evaluating the properties of a system, where quality characteristics are intended to match stakeholders' needs (e.g., functionality, performance, security, maintainability, etc.). However, as argued in [100], HSE drivers introduce new emerging needs, such as accountability, fairness, non-discrimination, transparency, and human autonomy. These needs call for a reconsideration of the traditional understanding of quality and quality standards, necessitating the inclusion of these new stakeholder requirements. This is evident in the recent guidelines and recommendations from institutional bodies (e.g., [2, 58, 93, 117]), which present principles and values for developing systems that are trustworthy and respectful of human rights and societal values.

  So far, some principles are being investigated more than others. For instance, while principles such as explainability, transparency, and fairness are receiving significant attention, values such as auditability, redress, and tradeoffs, which are related to accountability, have been overlooked. Additionally, these new quality dimensions embedding individual preferences and moral views raise questions about whether they should be classified as functional or non-functional requirements and whether such classification is meaningful.

  The research community has already started to explore these directions contributing with roadmaps (e.g., [76]), best practices (e.g., [75]), and frameworks (e.g., [12]) for responsible AI systems aiming at ensuring the responsible development and operation of such systems. However, more practical solutions are required to effectively influence the concrete development of modern digital systems.

- **RD2.2:** *Specification of humans, societal, and environmental requirements* – Building on the previous research direction regarding new quality dimensions, challenges emerge about how new and emerging stakeholders' needs are elicited and specified so that systems can interpret and leverage them in interactions with humans. An interesting proposal is presented by Townsend et al. [115], who introduced the concept of SLEEC (social, legal, ethical, empathetic, or cultural) rules aimed at facilitating the formulation, verification, and enforcement of the rules that AI-enabled and autonomous systems should follow. They outline a methodology for eliciting these rules, allowing domain experts, as well as stakeholders with different profiles, to formulate them in natural language. Subsequently, some approaches have been proposed to systematically translate these rules into a formal language to support automated reasoning, thus enabling their effective use in AI systems (e.g., [116, 128]). Nevertheless, more effort is needed to effectively apply them in real-world scenarios.

  Additionally, the ethical profile of the user can also evolve at runtime. We can envision an initial profiling approach that adopts a top-down strategy, establishing a default setting, which can then be complemented by a customization process that allows users to select subjective elements in accordance with their soft ethics. However, it is unrealistic to define a profile that accurately reflects the end user's preferences from the outset. Consequently, there is a need for a bottom-up refinement of the profile based on observable behaviors, perhaps utilizing an AI engine. Anyhow, defining an accurate ethical profile at design time is challenging because it is not straightforward for anyone to express their ethical views. This is where we encounter challenges, as a description of who we are, even a wrong one, becomes operational.

The work in [6, 99] investigates how to collect personal ethical preferences and their seamless and consistent combination with other relevant dimensions starting from a top-down ethical perspective. For instance, regarding the expression of HSE value, it is interesting to exploit the philosophical notion of dispositions [7, 9]. They are those properties that dispose towards other properties that manifest when some causal conditions are met. Transforming philosophical dispositions in logic specifications will make them readable by, e.g., autonomous decision-making systems, which can then quantify and compare different behavioral alternatives while accounting for HSE values. To transform the dispositional description into a logic specification, we can leverage the efforts made by the software engineering community to make writing logic specifications more manageable for practitioners through the use of patterns. In this direction, the work in [10] presents a single, comprehensive, and coherent pattern catalog collecting the existing qualitative, real-time, and probabilistic specification patterns. Following a similar approach, in [84, 85], specification patterns for robotic missions with support for quantitative properties are presented. Most probably, HSE values would require dedicated patterns. However, patterns, by definition, are derived from a body of knowledge from which they are identified, synthesized, and formulated. Nevertheless, we currently lack a clear body of knowledge regarding HSE properties, and it is not easy to get detailed and rich examples from experts. Additionally, the elicitation process includes stakeholders with diverse skills and cultural backgrounds who might not be used to work in such a structured manner.

To summarize, more work is needed in investigating and understanding how to specify HSE requirements and formulate them in a way that they can be programmatically accessed and used, e.g., by autonomous decision-making systems. The complexity comes also from bringing together different communities, like philosophers, computer scientists, and logicians towards new solutions that integrate concepts and research results from the different communities.

- **RD2.3:** *From tradeoff to runtime negotiation of requirements* – As highlighted in RD1.3, HSE values have a subjective dimension. This is also highlighted by the moral machine experiment of MIT: values collectively accepted are anchored to and depend on specific cultures and/or countries [11]. This implies that every single human has her own values that should be respected by the behavior of systems. However, in every single scenario of our daily life, we live in an environment that is shared with other humans and various technologies. Since humans are different from each other, there would be multiple and even conflicting values that should be guaranteed. This calls for a tradeoff and/or negotiation of values among the involved humans and digital entities [95]. We could be tempted to treat HSE values as traditional non-functional requirements like security and performance and their tradeoff, as done in [48]. In this research guideline, we argue that the traditional tradeoff, with strategies typically defined at design time, is not enough for the following reasons. First, we only identify the specific humans to be considered at runtime, along with their HSE profiles. Therefore, due to the high variability of humans and their inherent unpredictability, it is unfeasible to define tradeoff strategies at design time. Second, to find a solution that could satisfy the involved stakeholders, we would probably need a proper negotiation, where the involved humans should be guided and involved in identifying requirements that could be potentially relaxed or downgraded in order to find a solution that, to some extent, would satisfy all involved stakeholders. We are implicitly assuming that HSE requirements are "graduable", meaning that there is a degree to which the system satisfies the stated and implied needs of its various stakeholders [107].

## 5.3 Software Architecture and Design

The activities of software architecture and design are challenged by the increasing use of ML and generative AI. These technologies bring opportunities for automation never seen before. The way of producing code is changing, and it is becoming increasingly automated, with programmers shifting their activity from code writing to engineering the prompt or, more generally, in the specification of what the code should do. This subsection identifies four research directions concerning software architecture and design. RD3.1 warns of the need for embedding HSE values in the declarative specification of what the system should do. RD3.2 calls for new tactics, reference architecture, paradigms, and patterns for building digital systems for humanity. RD3.3 describes the research directions from the architectural and design point of view related to the subjectivity of systems behavior related to HSE drivers and, thus, to the need for adaptability and customizability of systems to evolving user profiles. Finally, RD3.4 introduces the concept and metaphor of human, social, and environmental debt, intended as the cost of not satisfactorily addressing these values during the engineering of systems.

- **RD3.1:** *Generative AI for humanity* – Generative AI is showing amazing performance in code generation, and this is revolutionizing the way software is produced. The design of systems will then be affected and changed by these technologies. On the one hand, we expect that the high-level and architectural design will be in the hands of humans since the main architectural decisions are of key importance for the system that will be built and for its alignment with business goals, technology drivers and, hopefully, also HSE drivers, as discussed in Section 3.3. On the other hand, lower-level design and software development are increasingly low-code and are substituted by generative AI. In a sense, this phase shifts from explicitly programming how the system should behave to providing a declarative specification of what the system should do. For instance, in LLMs, much attention is paid to the identification of strategies [68] and systematic and disciplined approaches to "prompt engineering" [103]. In order to make sure that the generated code will embed the HSE values, there is the need to consider these values during the declarative specification of what the system should do, e.g., in the prompt engineering of LLMs.
- **RD3.2:** *New tactics, reference architecture, paradigms, and patterns* – The popularity of ML has brought various challenges in architecting ML-enabled systems, both in terms of the pipeline that produces the ML model and the system that uses and interacts with the AI/ML components. While there has been some work aimed at collecting best practices in the software architecture design for ML systems [90, 122], this work is still in its infancy. There is a need to systematize a body of knowledge on the design of ML-intensive systems so that practitioners have better guidance in this complex domain. Moreover, architecture is tightly connected to quality: "Whether a system will be able to exhibit its desired (or required) quality attributes is substantially determined by its architecture" [16]. Since quality is extended to include HSE aspects, this extension impacts the architecture. Consequently, there is a need for new tactics, reference architecture, paradigms, and patterns tailored to the new quality needs.
- **RD3.3:** *Customizability at runtime based on evolving user profiles* – As discussed in RD1.3, the subjectivity of systems behavior related to HSE drivers is a source of variability that must be resolved at runtime. However, as also stated in RD2.2, we currently lack a clear body of knowledge regarding HSE properties, and it is not easy to gather such information into user profiles containing subjective elements that reflect those properties.
  Specifically, profiles can be profitably exploited to adjust the autonomy of AI-enabled systems in a way to redistribute the operational control among different parts of the system, as well as humans [87]. Adjustment can be made autonomously (self-adjustment) or by humans or external systems. The reasons for adjustable autonomy are various: (i) strategic choices since an autonomous system or a human can better perform in a specific task, (ii) ways to deal with the absence of trust and confidence of human users and then facilitate the human acceptance, (iii) ways to deal with the liability of a system, (iv) limitations or inability of an autonomous system in managing specific situations, etc. As surveyed in [87], the role of the human varies from teamwork, i.e., member of a non-hierarchical group [24], to supervised, where the human can supervise the autonomous system [129]. However, existing approaches to adjustable autonomy are static and lack flexibility in that autonomy is mostly adjusted according to predefined degrees or levels, and only a few of them take into account user preferences [40, 123, 129]. New approaches for adjustable autonomy, which are more flexible and tailored to HSE preferences, are needed with the main purpose of avoiding undesired consequences [73]. A paradigm shift from static approaches is specifically needed for ethical-aware and human-centric customization of adjustable autonomous systems to a groundbreaking dynamic one. That shift would support humans and social values and dynamically empower citizens in their interplay with the system. Two main research directions can be distinguished: (i) designing adjustable autonomous systems that provide user-centric customization of their autonomy, hence guaranteeing values such as privacy, fairness, and human dignity. This level of flexibility would empower the interaction between the user and autonomous systems, paving the way for a more user-friendly and ethical future of autonomous systems; (ii) developing software exoskeletons for adjusting the autonomy of autonomous systems to user-defined moral values while respecting privacy concerns. The user's ethical values drive the human-centric customization of the system autonomy and its subsequent (re-)distribution among the system's parts and actors. A tradeoff with privacy concerns might be required. For instance, the user might want to disclose personal information to gain some advantages only if the system's autonomy is flexible enough to allow her to decide when and how to destroy the disclosed information on its own.

- **RD3.4:** *Human, societal, and environmental debt* – Technical Debt (TD) has been largely studied for traditional systems [25, 43], and various tools have been proposed. However, there is still the need to understand technical debt in ML-intensive systems and adapt existing tools and techniques to analyze, track, and quantify design debt in traditional systems to the context of ML-intensive systems. Recently, researchers also proposed the dual concept of Technical Credit (TC) [64], which contrasts with the conventional TD-driven focus on the drawbacks of sub-optimal choices by aiming to characterize system features that can yield long-term benefits as the system evolves. Technical credit highlights the benefits of strategic investment and advocates for a paradigm shift toward recognizing the positive potential of forward-thinking. TC is a new concept that needs to be investigated in depth. Similarly to what was done in the past for social debt [113], the HSE drivers described in Section 3.3 trigger the investigation of a new metaphor: the HSE debt, intended as the cost for not addressing these values during the engineering of systems in a satisfactory way. When HSE becomes a business value, it will also become important to investigate the HSE credit to monitor and evaluate the investments done in the past to satisfy HSE values.

## 5.4 Verification and Validation

Despite the existence of amazing technologies that fascinate the world, we should admit that the global economy is extremely fragile in its dependence on computer systems. A proof of that was recently visible to everyone. A software update from a single cybersecurity company on July 2024, CrowdStrike, has affected computers running on Microsoft Windows. Since CrowdStrike is everywhere, e.g., numerous Fortune 500 companies use CrowdStrike's cybersecurity software to detect and block hacking threats, the software update caused disruptions and chaos in every single part of the world, affecting, e.g., hospitals, transportation, and hotels.[2] Companies slowly recovered from the outage since it required manual restarts of individual systems. From this unfortunate situation, we can learn various lessons. First, relying on a single company creates fragility in our technology ecosystem. This holds for small and big companies. Second, there is the need for verification and validation techniques even more than before. With the advent of AI, we lose control, and we need instruments to build robust systems at the service of humanity.

This subsection discusses these topics and presents three research directions. RD4.1 focuses on V&V for trustworthiness, highlighting the need for runtime techniques and the importance of considering also the new quality dimensions discussed in RD2.1. RD4.2 focuses on the importance of building trust and subjective acceptability of systems from the human perspective. RD4.3 focuses on the need for new instruments to support the continuous compliance with standards, regulations, or recommendations, not only for critical systems but also for systems that are not critical in the traditional understanding of criticality: as we learned from the recent story, a single problem in a non-critical system, like a problematic update, can affect the entire world. AI Act is starting to request a sort of certification for risky AI-based systems.

- **RD4.1:** *V&V for trustworthiness* – As discussed in Section 4.4, trustworthiness concerns the designing of systems in order to behave safely and guarantee security or quality aspects. Increasingly, systems are required to operate in semi-structured environments, often shared with humans, with partial knowledge about it, with limited controllability, and subject to various sources of uncertainty [28]. It becomes then impractical or even impossible to build guarantees of correct behavior of systems at design time: quality assurance techniques would be required to anticipate faulty scenarios, in spite of the complexities of runtime [27]. To mitigate the runtime complexity, V&V techniques make use of heuristics (e.g., test coverage [60]) and exploit the information available only at runtime [59]. In general, V&V try to move to runtime, with runtime verification techniques [14, 47], and field-based testing techniques [20, 106]. Despite the great efforts of the community, the need for V&V techniques that successfully work with the complexity of modern systems (like autonomous cars) is still not completely satisfied. This is exacerbated by the introduction of HSE values and their subjectivity dimension (see RD1.3). As explained in RD2.1, the new quality dimensions caused by new emerging and evolving stakeholders' needs call for new techniques to verify auditability, accountability, fairness, non-discrimination, transparency, etc. There is also the need for clear metrics to measure these different values, as well as techniques that guarantee a correct tradeoff and runtime negotiation of these new quality dimensions (see RD2.3).
- **RD4.2:** *Building trust* – With advancements in AI development, there is a growing need for a more nuanced understanding of trust in this technology [63]. AI is recognized as more than just a tool; its increasing integration into people's lives and the human-like traits it displays highlight its role as a socio-technical system dependent on human trust and adherence to ethical values. The interconnectedness of trust and ethics is explored in [32], where findings indicate that the ethical imperative for societal and environmental well-being is closely linked to human-like trust in AI. Additionally, accountability and technical robustness are identified as key factors influencing functionality trust in AI. In their study [55], the authors discuss the pervasive presence of AI in modern systems and the varying levels of acceptance it encounters across different fields. They highlight that some applications face skepticism from users, leading to low acceptance rates, while others see individuals excessively relying on AI recommendations. The authors argue that achieving an optimal balance of trust in AI requires careful calibration of expectations and capabilities. The work in [51] is motivated by the question: To what extent do people trust AI? Similarly to [55], the authors, by surveying 17.000 people from 17 countries, highlight that trust and acceptance vary depending on the specific AI application. For example, AI is generally regarded with greater trust in healthcare contexts compared to its use in human resource management. The authors highlight that there is robust global support for the principles of trustworthy AI. The majority of people consider these principles and their underlying practices essential for fostering trust. However, only a few believe that current governance, regulations, and laws are adequate to protect individuals and ensure the safe use of AI.

---

[2]Ironically, it happened a few days after the workshop colocated with FSE 2024 on the topics of this special issue. When discussing the future of software engineering, the majority of attendees were just fascinated by generative AI and LLMs, somehow leaving behind more traditional yet crucial software engineering techniques, decisive for building robust systems. More information on the software update and its consequences: https://edition.cnn.com/business/live-news/global-outage-intl-hnk/index.html.

Concerning the governance of AI, the study reveals that people have more confidence in national universities, research institutions, and defense organizations. Contrariwise, the least confidence is in governments and commercial organizations. Therefore, there is a need to build suitable and trustable governance institutes or associations. Recently, the AI advisory board of the United Nations recommended a holistic, inclusive, comprehensive, globally networked, agile, and flexible approach to governing AI for humanity as an instrument to "address the multifaceted and evolving challenges and opportunities AI presents on a global scale, [and, thus,] promoting international stability and equitable development" [21].

Concerning AI producers, it will become necessary to provide evidence of compliance with standards, regulations, best practices, and guidelines. This aspect is described in the following research direction.

- **RD4.3:** *Continuous compliance with standards, regulations or recommendations* – Due to the nature of ML-intensive systems and the need for continuous software updates after production, compliance with regulations, as well as standards and recommendations such as [31, 77, 120], must be maintained continuously, including during system operation and evolution. This poses various challenges in terms of techniques –e.g., there is the need for static analysis tools, for instance, to guarantee the compliance of code to best practice guidelines– but there is the need for processes that take into account also organizational issues and the relation with certification authorities [101]. Continuous compliance can be defined as the application of approaches and techniques to guarantee compliance with standards throughout a system's lifecycle, including post-production, when products are in use in the operational environment [101]. Also, it is not reasonable to assume that certification authorities can guarantee a "continuous" compliance process from one certification and the next. Instead, continuous compliance approaches and processes need to continuously (i) reassess and update artifacts and (ii) produce and store material and documentation that are relevant for the certification [101]. Much research and work is needed to make continuous compliance a consolidated, clear, and normal procedure. However, this is strongly required in various domains, like automotive with over-the-air updates to push new functionalities or new versions of software in the vehicles on the street, but also other domains like aerospace, which is, e.g., considering using AI on board of satellites [97]. Moreover, continuous compliance is also required by the *AI Act* [77], which, among other activities, requires companies to self-monitor compliance of their systems with the regulations. Probably, it will soon be required also in other parts of the world, with new regulations that will follow the so-called "Bruxelles effect" [23].

## 6 CONCLUSIONS

Digital systems are increasingly pervasive and part of our lives. Human beings are at risk, for, e.g., lack of transparency, unfair behavior, and biases of digital systems. To protect humans in their interplay with these systems, we believe that new engineering approaches are needed for the design and operation of digital systems for humanity. For this purpose, in this paper, we focused on the perspective of humans and their role in their co-existence with digital systems. Contrary to traditional development, which is driven mainly by business goals and technology drivers, we highlight the need to consider also human, social, and environmental values while engineering systems. This perspective adds further criticalities to known challenges, such as human-system interaction or trustworthiness, which were previously approached solely from a systems standpoint. The paper concludes with a concrete research roadmap of digital systems for humanity.

## ACKNOWLEDGEMENTS

This work has been partially funded by (a) the MUR (Italy) Department of Excellence 2023 - 2027, (b) the European HORIZON-KDT-JU research project MATISSE "Model-based engineering of Digital Twins for early verification and validation of Industrial Systems", HORIZON-KDT-JU-2023-2-RIA, Proposal number: 101140216-2, KDT232RIA_00017, (c) the PRIN project P2022RSW5W - RoboChor: Robot Choreography, (d) the PRIN project 2022JKA4SL - HALO: etHical-aware AdjustabLe autOnomous systems, and (e) the PRIN project 2022JAFATE - CAVIA: enabling the Cloud-to-Autonomous-Vehicles continuum for future Industrial Applications